\def\kb{\overline{k}}
\def\ob{\overline{\omega}}
\documentstyle[aps,prl,epsf,preprint,tighten]{revtex}
\begin{document}
\widetext
\title{Universal, finite temperature, crossover functions of the quantum
transition in the Ising chain in a transverse field
}
\author{Subir Sachdev}
\address{Department of Physics, P.O. Box 208120, Yale University,
New Haven, CT 06520-8120}
\date{September 13, 1995}
\maketitle

\widetext
\begin{abstract}
We consider finite temperature properties of the Ising chain in a transverse
field
in the vicinity of its zero temperature, second order quantum phase transition.
New universal crossover functions for static and dynamic correlators of the
``spin''
operator are obtained. The static results follow from an early lattice
computation
of McCoy, and a method of analytic continuation in the space of coupling
constants. The dynamic results are in the ``renormalized classical'' region and
follow
from a  proposed mapping of the quantum dynamics to the Glauber dynamics of a
classical
Ising chain.
\end{abstract}
\pacs{PACS:  }

\narrowtext
\section{Introduction}
\label{sec:intro}
\begin{figure}
\epsfxsize=6in
\centerline{\epsffile{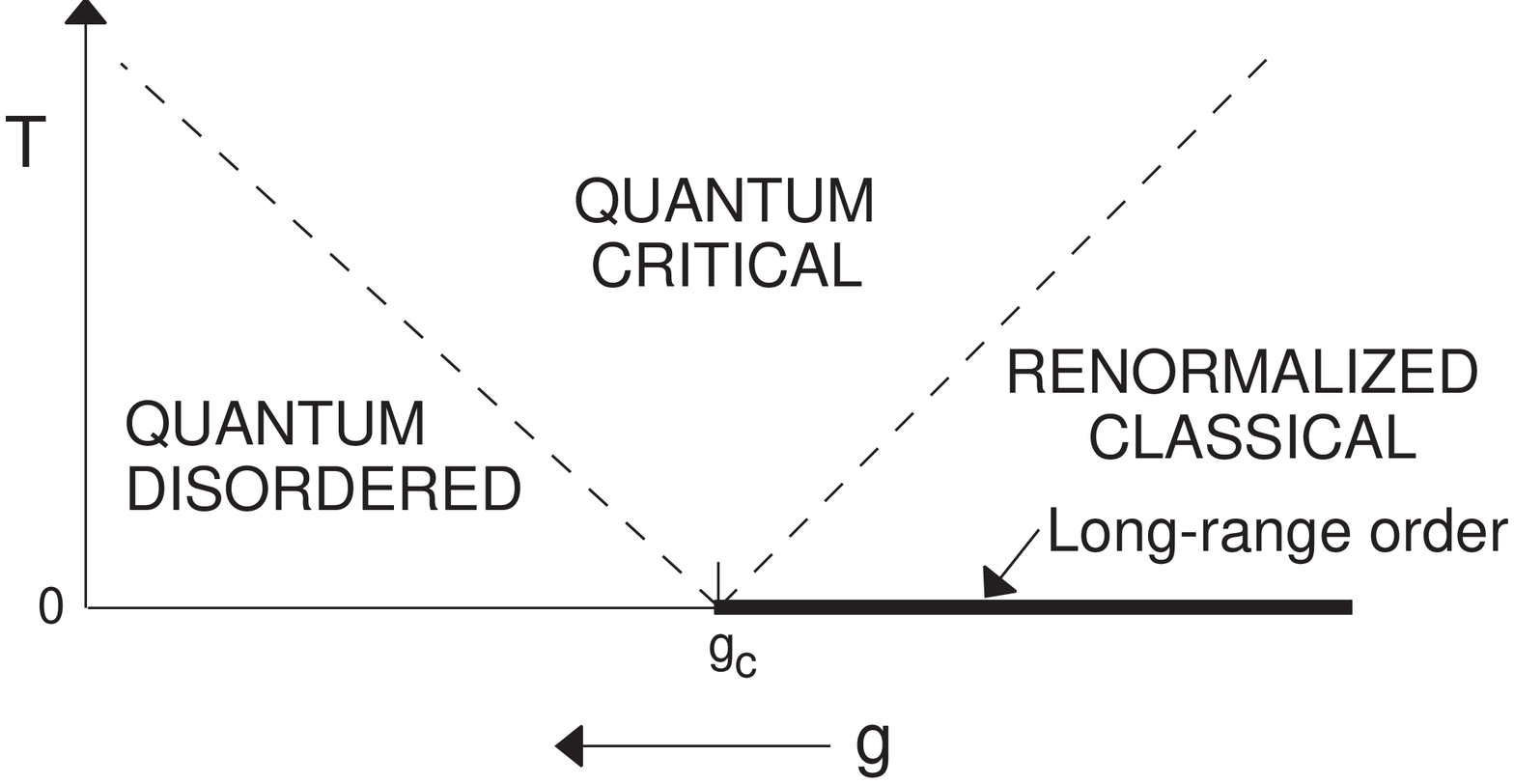}}
\vspace{0.5in}
\caption{Crossover phase diagram for the $d=1$ Ising chain ${\cal H}_I$ (Eqn.
(\protect\ref{hamising})). The quantum phase transition at $T=0$, $g=g_c$ has
exponents
$z=1$, $\nu = 1$. There is true long-range order at $T=0$ for $g < g_c$, as
indicated
by the thick line. The parameter $m \sim g_c - g$, so $m=0$ at the critical
point and
$m > 0$ on the ordered side. An essentially identical phase diagram applies to
the
$d=2$, $O(3)$ quantum rotor model~\protect\cite{CHN,CSY}, with the difference
that
the exponents $z=1$, $\nu \approx 0.7$; the crossover phase boundaries, which
occur
at $|g - g_c |^{z \nu} \sim T$, would then not be linear. All regions display
critical
fluctuations at short scales. In the renormalized-classical and
quantum-disordered regions,
these cross-over to behavior appropriate to the non-critical ground state
before thermal
effects have become apparent. In contrast, in the quantum-critical region, the
critical
fluctuations are quenched in a universal manner by thermal effects, and
characteristics
of the non-critical ground state do not appear at any scale~\protect\cite{CSY}.
}
\label{ising}
\end{figure}

The finite temperature properties of systems in the vicinity of
a zero temperature ($T$), second-order quantum phase transition have recently
seen
a great deal of attention~\cite{all,CHN,CSY,sss}.
In this paper we shall discuss properties of the quantum transition in
an Ising model in a transverse field in spatial dimension $d=1$. Apart from
its intrinsic interest as a possible experimental system,
the Ising model deserves scrutiny because it  has a finite
temperature phase diagram (see Fig~\ref{ising} and its caption) which is
extremely
similar to that of the
$d=2$,
$O(3)$ quantum rotor model~\cite{CHN,CSY}; the latter model is of relevance in
the
study of quantum Heisenberg antiferromagnets in $d=2$.
In Ref~\cite{CSY}, a number of results were obtained for universal scaling
functions of
the $d=2$, $O(N)$ rotor model by a $1/N$ expansion.  In the present paper, we
shall
describe the corresponding scaling structure for the $d=1$ quantum Ising model,
and
obtain exact results for crossover functions of some static and dynamic
observables.

An early study of the Ising model~\cite{kogut} obtained scaling results only at
$T=0$.
Scaling results at finite $T$, but with  transverse field set exactly at its
critical
value, can be obtained by conformal invariance (See {\em e.g.}
Ref~\cite{korepbook,statphys} and Section~\ref{dynamics}). Other, previous
finite $T$
results include corrections to scaling at the critical transverse
field~\cite{shrock},
and results at
$T=\infty$ which are beyond the scaling regime~\cite{perk,perk2}.

In this paper  we
shall obtain new static and dynamic scaling results for the Ising model at
finite
$T$, with the transverse field away from the critical point; the ground state
of the
model is therefore ``massive''.  The results are for two-point correlators of
the
``spin'' operator of the Ising model, and the static and dynamic results are
obtained
by rather different methods.\\
({\em i\/}) The static (or more precisely, equal time) results are obtained by
using
 a lattice model computation in an early paper by McCoy~\cite{mccoy}, and by a
method of analytic continuation in the space of coupling constants. A separate
scaling
analysis of McCoy's results was also considered earlier in  some limiting cases
 by Perk {\em
et. al.}~\cite{perk}, who, unlike us, did not consider crossover functions.
All but one of Perk {\em
et. al.}'s results agree with ours, and the disagreement is probably due to
a miscalculation in the last section of Ref~\cite{mccoy}; this issue will be
discussed
in more detail in Section~\ref{statics}.\\
({\em ii\/}) The dynamic results are not rigorous, but are based on a  physical
argument which suggests that in the ``renormalized-classical'' region
(Fig~\ref{ising}), the long-time, long-distance spin correlations
of the quantum Ising model should be described by
an effective classical model. We propose that the appropriate effective model
is
the Glauber dynamics~\cite{glauber} of a classical Ising chain. This proposal
then
allows us to compute dynamic scaling functions in the renormalized-classical
regime,
up to a single unknown damping constant.

The main technical innovation of this paper is the use of the method of
analytic
continuation in the space of coupling constants, which plays a central role in
the
computation of static correlations. This method was alluded to earlier by the
author,
Senthil and Shankar in Appendix C of a study~\cite{sss} of the dilute Bose gas
in
$d=1$; here we shall demonstrate its utility for the Ising model, and
provide more details on the methodology for both the Ising model and the Bose
gas. The
method relies on the fact that one-dimensional quantum systems with short-range
interactions do not have phase transitions at finite
$T$. As a result, all observable properties should be smooth functions of the
bare
coupling constants at any non-zero
$T$. There can be phase transitions at
$T=0$ however, separating two phases with very different physical
properties---see the
phase diagram and figure caption for the Ising model as function of coupling
constant
and $T$ in Fig~\ref{ising}. Raising $T$ slightly above the two $T=0$ phases,
gives
finite
$T$ states whose properties are again very different. However, the observables
in these
states must be related to each other by an analytic continuation in the
coupling
constant. This relationship becomes somewhat less counterintuitive when one
realizes
that the two states have to connect through the intermediate ``quantum-critical
region'' (see Fig~\ref{ising}) in which the observables only have a
smooth, subdominant, dependence on the coupling constant~\cite{CSY}. With the
method of
analytic continuation at hand, we can, from a knowledge of observables at low
$T$ above
one $T=0$ phase, deduce the observable at low $T$ above the critical point and
the other
$T=0$ phase.

As a by-product of our method of analytic continuation, we shall  obtain two
apparently
new integral representations of Glaisher's constant, $A_G$~\cite{wu}. This
constant
plays a central role in universal
amplitude ratios of two-dimensional field theories whose correlators  are
described by
Painlev\'{e} differential equations~\cite{korepbook,wu2,drouffe,andre}.
We do not have analytical proofs for our representations of $A_G$,  but will
instead
present numerical evidence which establish them beyond any reasonable doubt.
One integral representation will appear in the analysis of the Ising model
(Section~\ref{statics} and Appendix~\ref{cross}), while the other will be
deduced from
the Bose gas results of
Refs~\cite{sss,koreppapers} (Section~\ref{sec:bose} and
Appendix~\ref{boseproof}).

Our main new results on the the Ising chain in a transverse field will be
outlined
in Section~\ref{sec:ising}, with computational details being provided in
appendices~\ref{cross} and~\ref{glauberscale}; the static results will be
described in
Section~\ref{statics}, and dynamic results in Section~\ref{dynamics}. In
Section~\ref{sec:bose} we shall review earlier results on the properties of the
quantum
transition in the dilute Bose gas in
$d<2$~\cite{sss} and display exact results for
the universal crossover
functions of some static observables in $d=1$ as obtained in
Refs~\cite{korepbook,koreppapers,sss}; some new results will also be
established here,
although the main purpose of this section is to observe that
there are many strong formal
similarities between these Bose gas results and the new Ising model results of
this
paper. Section~\ref{conc} will discuss some general issues, including the
significance
of the similarity between the Bose gas and the Ising model.  This similarity
will
motivate a possible route to the solution of the remaining outstanding problem:
the
explicit  computation of dynamic correlations in the quantum Ising model (with
its
unitary Hamiltonian time evolution), without recourse to a mapping to the
phenomenological, dissipative,  Glauber model. Such a direct solution, and its
comparison to the phenomenological models, should increase our understanding of
the
relationship between the different approaches to dynamic phenomena in quantum
statistical system.

\section{Ising chain in a transverse field}
\label{sec:ising}
This section will describe our new results on
Ising chain in a transverse field. We will consider the Hamiltonian
\begin{equation}
H_I = -J \sum_{i} \left( g \sigma_x (i) + \sigma_z (i) \sigma_z (i+1) \right)
\label{hamising}
\end{equation}
where $J>0$ is an overall energy scale, $g>0$ is a dimensionless coupling
constant,
$\sigma_x (i)$, $\sigma_z (i)$ are Pauli matrices on a chain of sites $i$.
This model
has a $T=0$ quantum phase transition~\cite{kogut} at $g=g_c =1$ from a state
with
long-range-order with $\langle \sigma_z \rangle \neq 0$ ($g < g_c$), to a
gapped
quantum paramagnet ($g>g_c$). The critical exponents are
$z=1$ and $\nu=1$. The crossover phase diagram at finite $T$ is shown in
Fig~\ref{ising}. Notice the strong similarity between this and the phase
diagram for
the $d=2$ quantum rotor model~\cite{CHN,CSY}: in both cases there is long range
order
only at
$T=0$ for $g < g_c$, and the physical interpretation of the finite temperature
regions
is very similar (see the figure caption of Fig~\ref{ising}).
Further, at the smallest finite $T$
above the ordered phase, the spin correlation length, $\xi$, diverges
exponentially
with $T$:
\begin{equation}
 \ln \xi \sim (g_c - g)^{z\nu} /
T.
\end{equation}
In the $d=2$ quantum rotor model this behavior of $\xi$ is due to the
properties of the
$d=2$ classical non-linear sigma model~\cite{CHN}; in the $d=1$ Ising model,
the
correlation length is of order the mean spacing between the thermally
activated kink
and anti-kink excitations, which have an energy $\sim (g_c - g)^{z\nu}$.

We consider, now, the universal functions describing the crossovers in
Fig~\ref{ising} for $H_I$. First, it is necessary to define the appropriate
renormalized couplings. It is known that in the vicinity of the critical point,
the
Ising model is described by a continuum field theory of free Majorana
fermions~\cite{drouffe}. We choose the mass $m$  and the velocity
$c$ of the Majorana fermions as our renormalized couplings  (the energy of a
fermion
with wavevector
$k$  is therefore $\left( m^2 c^4 + \hbar^2 c^2 k^2 \right)^{1/2}$); notice
that
these renormalized couplings are defined at $T=0$.  The relationship between
$m$ and
$c$, and the bare couplings of the lattice model  is non-universal and, in
general,
difficult to determine; for the special case of $H_I$, which happens to be
integrable,
we have
\begin{equation}
m = \frac{\hbar^2 (g_c-g)}{2 J a^2}~~\mbox{and}~~c = \frac{2Ja}{\hbar},
\end{equation}
where $a$  is
the lattice spacing. The Majorana fermion mass $m$ can have either sign, and
moving $m$
through 0 tunes the system across the transition; we have
chosen $m > 0$ to correspond to the ordered side.  Scaling would suggest that
$|m|c^2
\sim |g-g_c|^{z\nu}$, and the value of $m$ for $H_I$
agrees with $z\nu = 1$. The velocity, $c$, simply sets the relative distance
and time
scales and may be assumed to be constant across the transition.

As we discussed in Section~\ref{sec:intro},
there can be no phase transitions in this Ising model at any non-zero $T$.
We may therefore conclude that all properties of
$H_I$ are analytic functions of the bare coupling $g$ at all non-zero $T$.
Accidentally, because $z \nu = 1$, analyticity in $g$ is equivalent to
analyticity in
$m$. This is a powerful principle which we shall use below.

Universality appears in the continuum limit $a \rightarrow 0$ at fixed $m$,
$c$,
and $T$ (this requires $J\rightarrow \infty$ and $g \rightarrow g_c$). An
equivalent
``condensed matter'' perspective is that we are studying length scales much
larger than $a$,
energy scales much lower than $J$, but of order scales that can made out of
combinations of
$m$, $c$, and $T$ . Following
arguments in Ref~\cite{CSY}, we may conclude that, in this limit
\begin{equation}
\left\langle \sigma_z (x, t) \sigma_z (0,0) \right\rangle
= Z (k_B T)^{1/4} \Phi_I \left(
\frac{k_B T}{\hbar c} x , \frac{k_B T}{\hbar} t, \frac{mc^2}{k_B T} \right),
\label{isingscale}
\end{equation}
where $Z$ is a non-universal normalization constant, and $\Phi_I$ is a
fully universal crossover function; the above scaling form reproduces the
$T=0$ results in the limit
$T \searrow 0$.
The non-universality of
$Z$ is linked with the  anomalous dimension of the $\sigma_z$ operator,
indicated by
the
$T^{1/4}$ prefactor.
This non-universality also means that a single normalization condition  is
necessary to
set the overall scale of
$\Phi_I$: this will be specified in the next subsection
and leads to the value
\begin{equation}
Z = \frac{1}{J^{1/4}}
\end{equation}
for the specific lattice Hamiltonian $H_I$.

We emphasize that we are studying here correlators of  the `spin' operator
$\sigma_z$.
The expression for $\sigma_z$ in terms of  Majorana  fermions involves an
infinite
string of fermion operators~\cite{drouffe}. In contrast, the `energy' operator
$\sigma_x$ is a bilinear in the Majorana fermions; its scaling properties are
much
simpler and easily computable~\cite{drouffe}.

\subsection{Statics}
\label{statics}
This section will present our new results for the crossover functions of the
two-point,
equal time, correlators of $\sigma_z$. In particular, the long-distance limit
of this correlator has the form
\begin{equation}
 \left\langle
\sigma_z (|x|\rightarrow \infty, 0)
\sigma_z (0,0)
\right\rangle \sim A_I \exp(-|x|/\xi_I),
\end{equation}
where $A_I$ is an amplitude, and $\xi_I$ is the correlation length. The result
(\ref{isingscale}) implies that these must obey the scaling forms
\begin{eqnarray}
\xi_I^{-1} &=& \frac{k_B T}{\hbar c} f_I \left( \frac{mc^2}{k_B T} \right)
\nonumber \\
A_I &=& Z (k_B T)^{1/4} g_I \left( \frac{mc^2}{k_B T} \right),
\label{isingcross1}
\end{eqnarray}
where $f_I (s)$ and $g_I (s)$ are completely universal functions of $-\infty <
(s =
mc^2 /k_B T) < \infty$.
One of our main new results is the
explicit, closed-form, expression for these universal functions:
\begin{eqnarray}
f_I (s) &=& \int_0^{\infty} \frac{dy}{\pi} \ln \coth \frac{\sqrt{y^2 + s^2}}{2}
+
|s| \theta(-s) \nonumber \\
\ln g_I (s) &=&    \int_{s}^{1} \frac{dy}{y} \left[ \left(
\frac{df_I(y)}{dy} \right)^2 - \frac{1}{4} \right] + \int_{1}^{\infty}
\frac{dy}{y}
\left(\frac{df_I(y)}{dy} \right)^2,
\label{isingcross2}
\end{eqnarray}
where $s$ can take any finite value between $\pm \infty$, and $\theta$ is the
unit step
function
\begin{equation}
\theta(y) = \left\{ \begin{array}{cc} 1 & y > 0 \\
0 & y < 0 \end{array} \right..
\end{equation}
Details of the computation are provided in Appendix~\ref{cross}.
We have set the overall scale of $g_I$, and hence that of $\Phi_I$, by the
normalization condition
\begin{equation}
\lim_{s \rightarrow +\infty} \frac{g_I (s)}{s^{1/4}}
= 1.
\end{equation}

A key property of the crossover functions
$f_I(s)$ and $g_I(s)$ is that they are analytic for all $-\infty < s <\infty$.
The
apparent singularity at $s=0$ in the definition of $f_I (s)$ in
(\ref{isingcross2}) is deceptive: it is
not difficult to show that all derivatives of $f_I(s)$ are actually
finite at $s=0$ (see Appendix~\ref{cross}). The analyticity of $g_I (s)$
follows
from its definition in (\ref{isingcross2}), the analyticity of $f_I (s)$ and
the fact
that
\begin{equation}
\left. \frac{df_I}{ds} \right|_{s=0} = - \frac{1}{2}.
\end{equation}
This last result can be obtained by an elementary quadrature, and ensures that
there
there is no singularity in the integration at $y=0$ in the definition of $g_I
(s)$ in
(\ref{isingcross2}).
Indeed, these analyticity properties were an important ingredient in the
derivation of
(\ref{isingcross2}) in Appendix~\ref{cross}; there we take the continuum limit
 of a lattice model result
of McCoy~\cite{mccoy} for $s > 0$, and obtain results for $s \leq 0$  by
analytic continuation in $s$.

\begin{figure}
\epsfxsize=5in
\centerline{\epsffile{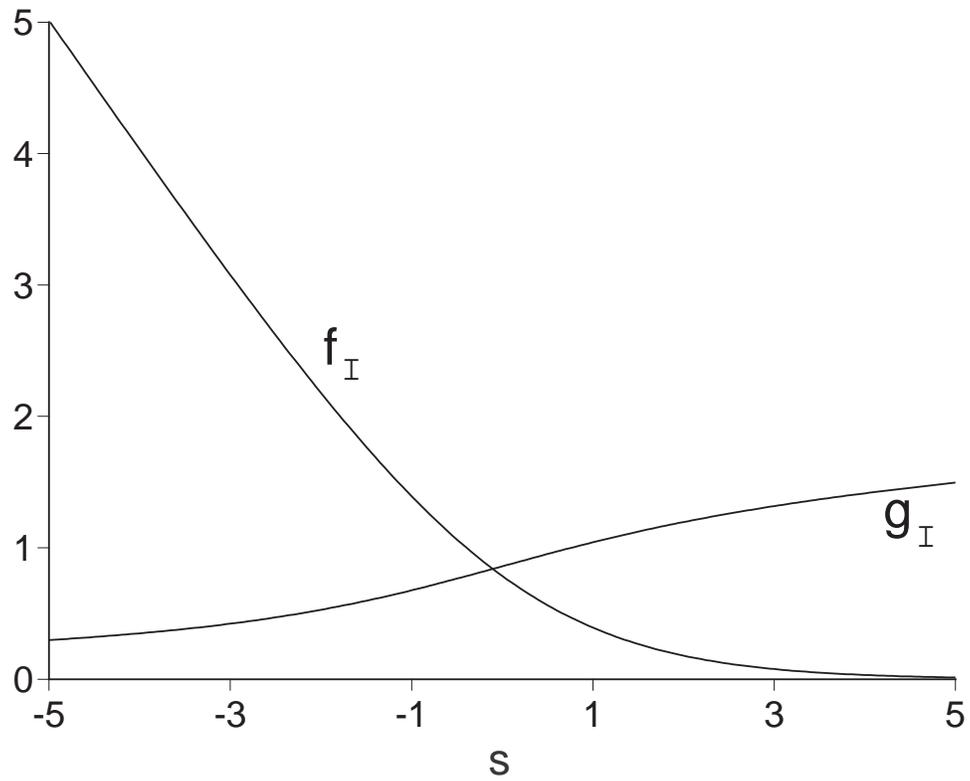}}
\vspace{0.5in}
\caption{The universal crossover functions $f_I$ and $g_I$ of the Ising model
as a function of $s = m c^2/k_B T$}
\label{fg}
\end{figure}

We show in Fig~\ref{fg} a plot of the functions $f_I (s)$, $g_I (s)$.
Some asymptotic limits can also be obtained by elementary means
from (\ref{isingcross2})
\begin{equation}
f_I (s) = \left\{
\begin{array}{ccc}
\sqrt{2 s/\pi}e^{-s}  & ~~ & s \rightarrow +\infty \\
\pi/4 & ~~ & s=0 \\
|s| + \sqrt{2 |s|/\pi}e^{-|s|}  & ~~ & s \rightarrow -\infty \end{array}
\right.
\label{asymptote1}
\end{equation}
The corresponding asymptotic limits for $g_I (s)$ are not so simple.
In Appendix~\ref{cross} we find
\begin{equation}
g_I (s) = \left\{
\begin{array}{ccc}
s^{1/4} & ~~ & s \rightarrow +\infty \\
{\cal C}_1  & ~~ & s=0 \\
|s|^{-3/4}  & ~~ & s \rightarrow -\infty \end{array}
\right.
\label{asymptote2}
\end{equation}
The universal number ${\cal C}_1$ was computed numerically
(Appendix~\ref{cross}),
and we found
\begin{equation}
{\cal C}_1 = \pi^{1/4} e^{1/4} 2^{1/12} A_G^{-3} + \epsilon,
\end{equation}
where $A_G$ is Glaisher's constant (given explicitly by
$\ln A_G = 1/12 - \zeta^{\prime} (-1) $ where $\zeta(x)$ is the Riemann
zeta function).
We have
numerically established that $|\epsilon| < 10^{-20}$, and there is essentially
no doubt
that in fact $\epsilon=0$; if so, we have
\begin{equation}
\ln \left( \pi^{1/4} e^{1/4} 2^{1/12} A_G^{-3} \right ) =
\int_{0}^{1} \frac{dy}{y} \left[ \left(
\frac{df_I(y)}{dy} \right)^2 - \frac{1}{4} \right] + \int_{1}^{\infty}
\frac{dy}{y}
\left(\frac{df_I(y)}{dy} \right)^2,
\end{equation}
as one of our new integral representations of $A_G$. We obtained our proposed
value
of ${\cal C}_1$ by using a conformal invariance argument to connect to a
$T=0$ result of Pfeuty~\cite{kogut} (see Eqn (\ref{conform}) below).

Perk {\em et. al.}~\cite{perk} have also studied the continuum limit of
McCoy's results~\cite{mccoy} for limiting regimes at low
$T$, although they did not obtain crossover functions. The results in
(\ref{asymptote1},\ref{asymptote2}) are in  agreement with theirs, except that
their
value for
$A_I$ at
$s=0$ disagrees with our value for ${\cal C}_1$.
The discrepancy at the critical point, $m=0$, between their
value for the amplitude $A_I$  at low $T$ and
Pfeuty's~\cite{kogut} result at $T=0$, led Perk {\em et. al.}~\cite{perk} to
suggest that the crossover function connecting to the $T=0$ result (Eqn
(\ref{conform}) below)
must break down. Our results show that there  is nothing wrong with the
crossover
function in (\ref{conform}) (indeed, it is demanded  by conformal invariance),
but it
is the value of $A_I$ at $s=0$ in Ref~\cite{perk} which is incorrect.
The error appears to be traceable to an earlier error in
the Section 6 of McCoy's paper~\cite{mccoy} which computed correlations at
$m=0$
and finite $T$. In the present paper we have only used McCoy's results for
$m>0$
from his Section 3, and have verified his computations in this case; for
completeness,
Appendix~\ref{cross} contains a sketch of the required intermediate steps.

\subsection{Dynamics}
\label{dynamics}

\begin{figure}
\epsfxsize=6in
\centerline{\epsffile{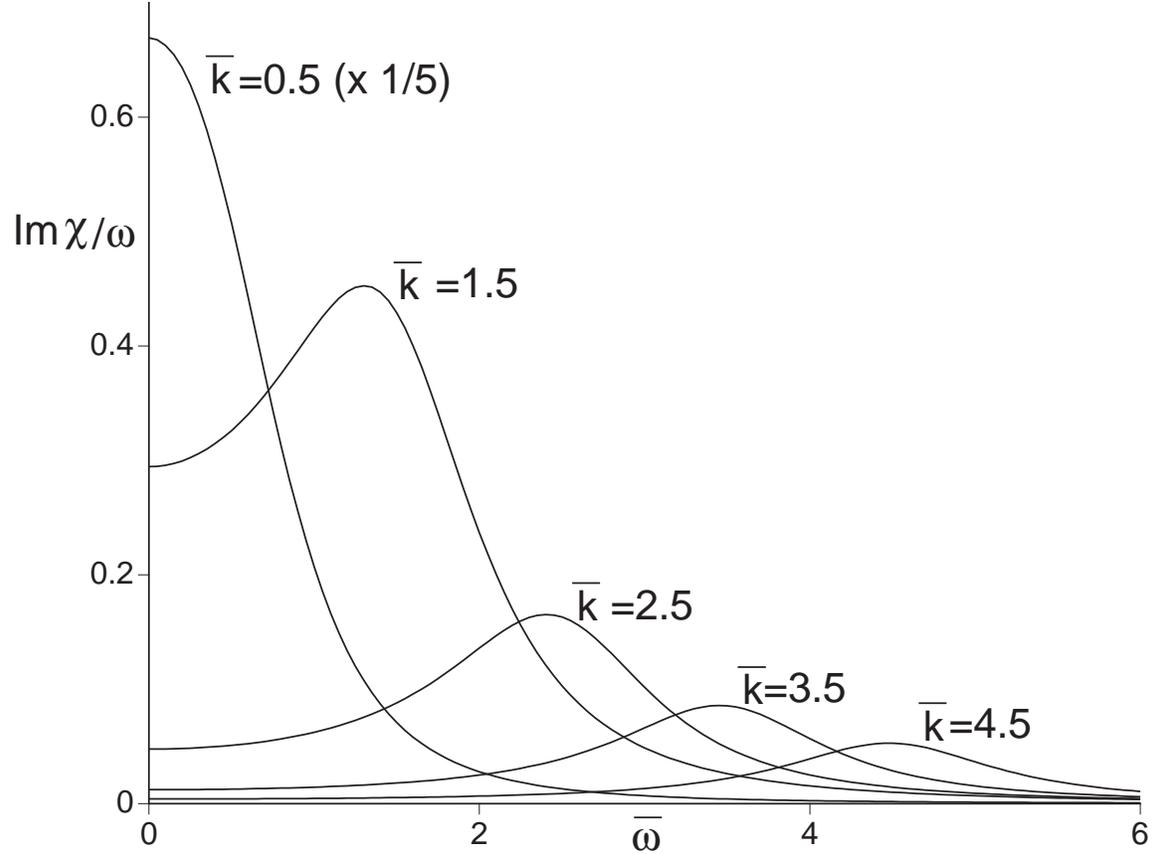}}
\vspace{0.5in}
\caption{
The spectral density of the dynamic susceptibility,  $\mbox{Im} \chi (k,
\omega) /
\omega$, of the transverse field Ising model in $d=1$ at its critical coupling
$g=g_c$
($m=0$) but with $T$ finite. The scale of the vertical axis is determined by
the
non-universal constant $Z$ in  (\protect\ref{chiscale}), and has been chosen
arbitrarily. We have
 $\kb=\hbar ck/k_B T$, $\ob=\hbar
\omega/k_B T$; the horizontal axis is therefore measured in dimensionless units
and its scale
is meaningful.}
\label{chiplot}
\end{figure}

We begin by reviewing and remarking on some earlier results for dynamic
correlations.\\
({\em i\/}) The scaling function $\Phi_I$ can be
obtained exactly at $m=0$ by using the conformal invariance of the
$T=0$ critical theory. One
finds~\cite{luther,perk,cardy,sss,korepbook,statphys}
in imaginary time:
\begin{equation}
\Phi_I ( \overline{x}, -i \overline{\tau}, 0) = \frac{2^{-1/8} g_I (0)}{
\left( \cosh(2 \pi \overline{x}) - \cos (2 \pi \overline{\tau}) \right)^{1/8}},
\label{conform}
\end{equation}
where we are using $\overline{x} = (k_B T/\hbar c) x$ and $\overline{\tau} =
(k_B T/\hbar) \tau$.
Note that
\begin{equation}
\Phi_I ( |\overline{x}| \rightarrow 0,0 ,0) = \frac{g_I (0)}{ |2 \pi
\overline{x}|^{1/4}},
\end{equation}
 which agrees with the $T=0$ result of Ref~\cite{kogut}, while
\begin{equation}
\Phi_I (|\overline{x}| \rightarrow \infty,0,0) = g_I (0) e^{-\pi
|\overline{x}|/4},
\end{equation}
which agrees with (\ref{isingcross2},\ref{asymptote2})); this is the
agreement between the low $T$ and $T=0$ results claimed earlier. We can also
Fourier
transform (\ref{conform}) and analytically continue to real frequencies to
obtain the
dynamic susceptibility $\chi (k, \omega )$, which is the Fourier transform of
the
retarded
$\sigma_z$, $\sigma_z$ commutator. There are a number of interesting subtleties
that
appear in this process---see Refs~\cite{sss,statphys};  here we display the
final
result:
\begin{equation}
\chi(k, \omega ) =
\frac{Z g_I (0) \Gamma (7/8)}{4 \pi \Gamma (1/8)} \frac{\hbar c}{(k_B T)^{7/4}}
\frac{\displaystyle \Gamma \left( \frac{1}{16} + i \frac{\ob + \kb}{4 \pi}
\right)
\Gamma \left( \frac{1}{16} + i \frac{\ob - \kb}{4 \pi} \right)}
{\displaystyle \Gamma \left( \frac{15}{16} + i \frac{\ob + \kb}{4 \pi} \right)
\Gamma \left( \frac{15}{16} + i \frac{\ob - \kb}{4 \pi} \right)}.
\label{chiscale}
\end{equation}
We have used $\kb = \hbar c k/k_B T$ and $\ob = \hbar \omega/k_B T$;
recall also that this result is valid only at the critical coupling $m=0$,
$g=g_c$.
A plot of the spectral density $\mbox{Im} \chi (k, \omega )/\omega$ as a
function of $\omega$
is shown
in Fig~\ref{chiplot} for a number of values of $\kb$. Notice that there is
relaxational behavior
for small $\kb$ ({\em e.g.} $\kb = 0.5$) with a peak in the spectral density at
$\ob =0$;
at large $\kb$ the peak moves to $\ob \approx \kb$ indicating a crossover to
the critical
excitations of the $T=0$ ground state. \\
({\em ii\/}) Perk {\em et.al.}~\cite{perk} obtained a set of non-linear,
partial
differential equations satisfied by an observable related to $\Phi_I$. However,
$T$
does not appear as a parameter in any of the equations, which are therefore
identical
to those obtained at $T=0$ (see Ref~\cite{drouffe} and references therein). All
effects
of temperature appear in the initial conditions, which presumably have to be
specified
on some space-like surface.  At the critical transverse field, Perk {\em et.
al.}~\cite{perk} used the differential equations to conjecture the form
(\ref{conform}), apart from the overall scale (recall the discussion in
Section~\ref{statics} on problems in Ref~\cite{perk} with the overall scale).
It
remains to be seen whether this approach can yield results on $\Phi_I$ away
from
criticality.\\
({\em iii\/})A  number of studies have examined finite $T$, time-dependent,
on-site
($x=0$) correlations of the lattice model at the critical coupling
$g=g_c$~\cite{shrock}. The scaling limit of all of these results is contained
already
in (\ref{conform}); instead, they contain a great deal of additional
information on the
non-universal corrections to scaling in the nearest-neighbor model
$H_I$.\\
({\em iv\/}) We also recall results of Perk {\em et.al.}~\cite{perk,perk2} for
time-dependent correlation functions at $T=\infty$ for the lattice model. These
results
are dominated entirely by non-universal lattice effects, and contain no overlap
with
our universal scaling form (\ref{isingscale}).

The major gap in our knowledge of time-dependent correlations is the form of
the
dynamic scaling function away from the critical-coupling ($m \neq 0$). In
Section~\ref{conc}, we will sketch a strategy by which these may be obtained.
In the remainder of this section we will use some phenomenological arguments to
obtain
some new results on the form of the dynamic scaling functions
in the renormalized-classical region of Fig.~\ref{ising}.

Notice that in the renormalized-classical (RC) region ($mc^2 \gg k_B T$),  the
correlation length becomes (from (\ref{isingcross2}), (\ref{asymptote1}) and
also
Ref~\cite{perk}) exponentially large:
\begin{equation}
\xi_{RC} =  \left( \frac{\pi \hbar^2}{2 m k_B T} \right)^{1/2}  \exp\left(
\frac{mc^2}{k_B T}
\right)
\end{equation}
In particular, $\xi_{RC}$ is significantly larger than the  length scale $\sim
\hbar
/mc$ at which critical fluctuations are quenched, and the ordering of the $T=0$
ground state appears. Using arguments similar to those for the $d=2$ rotor
model~\cite{CHN}, we may conclude that at  length scales $\gg \hbar /m c$,  but
of
order $\xi_{RC}$, an effective {\em classical} model of the system should be
adequate.
We propose here that the appropriate classical model is the Glauber dynamics of
the
$d=1$ classical Ising chain~\cite{glauber}. Formally, the above arguments
mean~\cite{CSY} that in the RC region, the  scaling function, $\Phi_I$, of 3
arguments in
(\ref{isingscale})  collapses into a reduced scaling function, $\phi_{RC}$, of
2
arguments:
\begin{equation}
\left\langle \sigma_z (x, t) \sigma_z (0,0) \right\rangle
= N_0^2 \phi_{RC} \left(
 \frac{x}{\xi_{RC}} , \gamma \frac{\hbar t}{m \xi_{RC}^2}\right)
{}~~~~~~~mc^2 \gg k_B T, |x| \gg
\frac{\hbar }{mc}, |t| \gg \frac{\hbar}{mc^2},
\label{RCscale}
\end{equation}
where $N_0 = Z^{1/2} (m c^2)^{1/8} = \left\langle \sigma_z
\right\rangle_{T=0}$ is the ground state magnetization,
and $\phi_{RC}$ is precisely the
dynamic scaling function of the
low $T$ limit of the Glauber model~\cite{glauber} (a
very similar collapse occurs in the Luttinger liquid region of the dilute Bose
gas; see Sec~\ref{sec:bose} and Ref.~\cite{sss}). Notice that the time $t$
scales
as $\xi_{RC}^2$---this is because the Glauber model has dynamic exponent $z=2$.
The dimensionless constant $\gamma$ determines the relaxation rate constant of
the
Glauber model; on
dimensional grounds we expect
\begin{equation}
\gamma = {\cal R} \left(\frac{k_B T}{m c^2}\right)^{\rho}
\end{equation}
where ${\cal R}$ is a
universal number, and
$\rho$ an unknown exponent.
The functional form of the universal function
$\phi_{RC}$ can be determined by taking the scaling limit of Glauber's
solution~\cite{glauber} at low $T$; we obtained (see
Appendix~\ref{glauberscale}):
\begin{equation}
\phi_{RC} (\tilde{r}, \tilde{t} ) = \frac{e^{\tilde{r}}}{2} \mbox{erfc} \left(
\sqrt{\frac{|\tilde{t}|}{2}} + \frac{\tilde{r}}{\sqrt{2|\tilde{t}|}} \right) +
\frac{e^{-\tilde{r}}}{2} \mbox{erfc} \left(
\sqrt{\frac{|\tilde{t}|}{2}} - \frac{\tilde{r}}{\sqrt{2|\tilde{t}|}} \right),
\label{erfc}
\end{equation}
where $\mbox{erfc}$ is the complementary error function. Notice that $\phi_{RC}
(\tilde{r}, 0) = e^{-|\tilde{r}|}$; combined with (\ref{RCscale}), this result
agrees
with the $m c^2 \gg k_B T$ limit of (\ref{isingscale})-(\ref{asymptote2}).

\section{Dilute Bose gas}
\label{sec:bose}
In this section we will review earlier results, and obtain some new results, on
the
finite $T$ crossovers in the dilute Bose gas in $d<2$. There are two main
reasons for
doing this here:\\
({\em i\/}) We will show that the results for the crossover functions have
strong
formal similarities to the corresponding quantities for the Ising model. In
Section~\ref{conc}, this similarity will motivate proposals for
computing dynamic correlations directly in the quantum Ising model.\\
({\em ii\/}) There are two distinct forms for the crossover functions in the
literature---one obtain by Korepin {\em et.
al.\/}~\cite{korepbook,koreppapers},
and other in Ref.~\cite{sss} by the method of analytic continuation. In this
section,
and Appendix~\ref{boseproof}, we will establish that these two forms are in
fact equal,
and in the process obtain a second new integral representation of Glaisher's
constant.

\begin{figure}
\epsfxsize=6in
\centerline{\epsffile{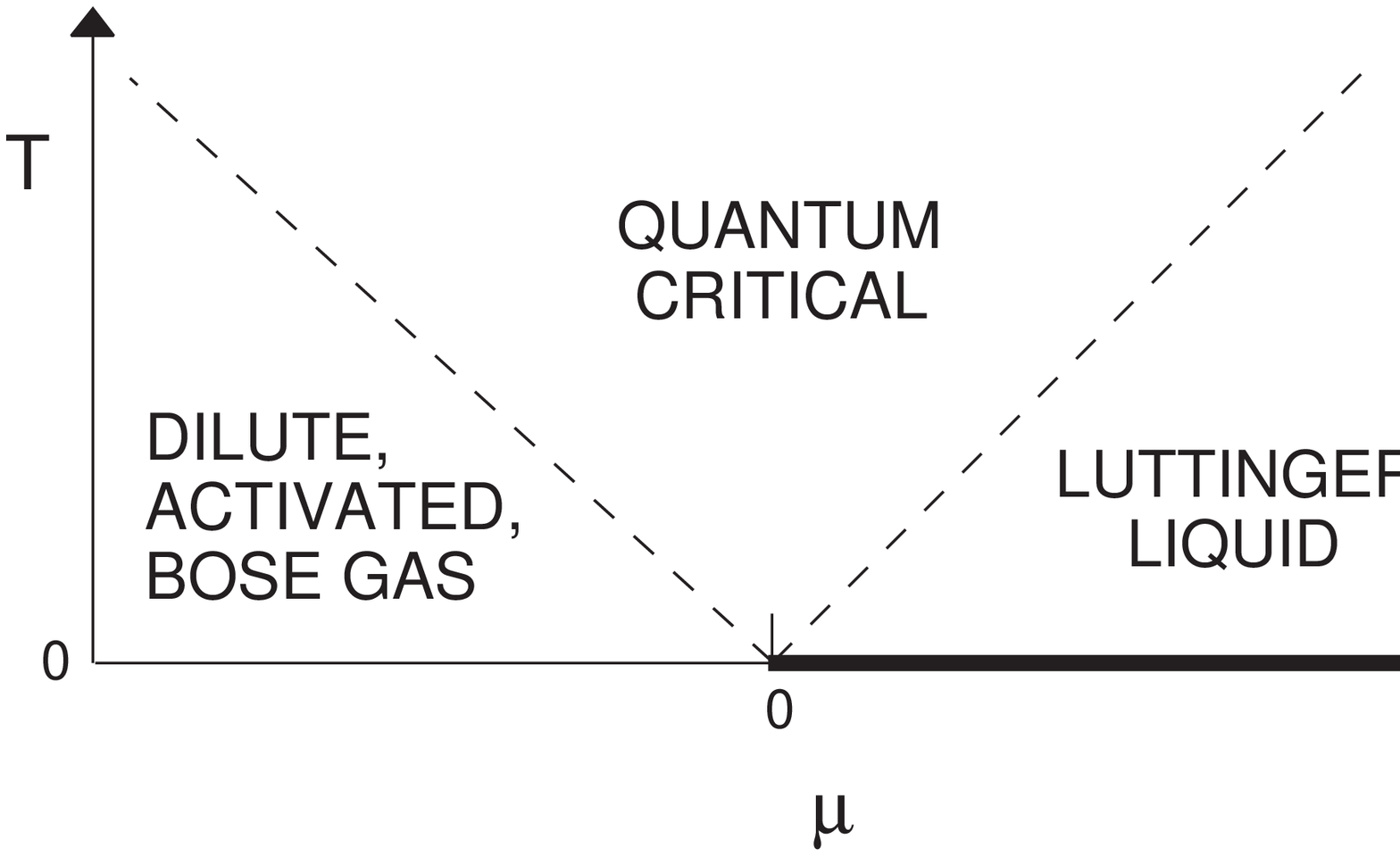}}
\vspace{0.5in}
\caption{Crossover phase diagram for a dilute Bose gas  in $d=1$ from
Ref~\protect\cite{sss}. The $T=0$ quantum phase transition occurs at chemical
potential
$\mu=0$ and has exponents $z=2$,
$\nu =1/2$.
The temperature $T$ is the most important energy scale in the quantum-critical
region.
The Luttinger liquid has power-law correlations at $T=0$ (indicated by the
thick line),
while there is exponential decay at any finite $T$. The same phase diagram
applies for
$1 < d < 2$ except that there is true long-range order at $T=0$ on the thick
line,
and the nomenclature ``incipient superfluid'' is more  appropriate than
``Luttinger
liquid''.}
\label{bose}
\end{figure}

We begin by reviewing known results for the Bose gas. Consider a dilute gas of
non-relativistic, spinless, bosons of mass $m_B$, at a chemical potential
$\mu$, in $d$ spatial
dimensions. The partition function of the system is given by
\begin{displaymath}
Z = \int {\cal D} \Psi \exp\left( -\frac{1}{\hbar} \int_0^{\hbar/k_B T}
d \tau {\cal L}(\tau) \right)
\end{displaymath}
\begin{eqnarray}
{\cal L}(\tau) = \int d^d &x& \left[
\hbar \Psi^{\dagger} (x,\tau) \frac{\partial \Psi (x,\tau)}{\partial \tau}
- \frac{\hbar^2}{2m_B} \Psi^{\dagger} (x,\tau)\nabla^2 \Psi (x,\tau) -
\mu |\Psi (x,\tau) |^2\right]\nonumber\\
&~&~~~~~~~~~~~~~~~~~~~~~~~~~~~~~~~~
+ \frac{1}{2} \int d^d x d^d x^{\prime} |\Psi (x,\tau) |^2 v(x-x^{\prime})
|\Psi (x^{\prime}, \tau )|^2 ,
\label{coherent}
\end{eqnarray}
where $\Psi(x,\tau)$ is a canonical Bose field over spacetime $x,\tau$.
The bosons experience a repulsive, short-range, but otherwise arbitrary, mutual
interaction, $v(x)$. At $T=0$, there is
a continuous
quantum phase transition as a function of $\mu$ at $\mu=0$, when the density
of particles in the
ground state changes from zero to a finite value. This transition has  dynamic
exponent
$z=2$ and correlation length exponent $\nu = 1/2$~\cite{fisher}. The finite $T$
crossovers near this quantum critical point are especially universal for $d<2$
(in the
sense that everything, including the scale factors, is independent of the
details of the interaction
$v(x)$), and are summarized in the phase diagram in Fig~\ref{bose}~\cite{sss}.
In the
vicinity of the critical point, the two-point correlator of $\Psi (x,t)$, where
 now
$t$ is real time, obeys~\cite{sss}
\begin{equation}
\left\langle \Psi(x, t) \Psi^{\dagger} (0,0) \right\rangle
= \left( \frac{2 m_B k_B T}{\hbar^2} \right)^{d/2} \Phi_B \left(
\frac{\sqrt{2 m_B k_B T}}{\hbar} x , \frac{k_B T}{\hbar} t, \frac{\mu}{k_B T}
\right)
\label{zeroscale}
\end{equation}
with $\Phi_B$ a completely universal scaling function, independent of the boson
interaction potential. Note that, unlike the Ising model, there are no
anomalous
dimensions and it was not necessary to define any renormalized couplings. Only
{\em
bare} coupling constants appear in the scaling form: this ``no-scale-factor
universality'' is a consequence of the  fluctuationless ground state  for $\mu
\leq 0$,
and the ultraviolet finiteness of the critical quantum field theory~\cite{sss}.
 These
last properties are rather special and, in particular, are not present in the
Ising
model considered below. In $d=1$, closed-form expressions  for crossover
functions,
corresponding to some limiting cases of
(\ref{zeroscale}), have been obtained~\cite{korepbook,koreppapers,sss}. The
long-distance limit of the  equal time correlator is argued to have the leading
behavior
\begin{equation}
 \left\langle
\Psi(|x|\rightarrow \infty, 0)
\Psi^{\dagger} (0,0)
\right\rangle \sim A_B \exp(-|x|/\xi_B);
\end{equation}
{}From (\ref{zeroscale}) we see that the amplitude $A_B$ and the correlation
length
$\xi_B$ must satisfy the scaling forms
\begin{eqnarray}
\xi_B^{-1} &=& \frac{\sqrt{2 m_B k_B T}}{\hbar} f_B \left( \frac{\mu}{k_B T}
\right)
\nonumber \\
A_B &=& \frac{\sqrt{2 m_B k_B T}}{\hbar} g_B \left( \frac{\mu}{k_B T} \right).
\label{bosecross1}
\end{eqnarray}
The crossover functions, $f_B$, $g_B$, in the form obtained in Ref~\cite{sss}
are
\begin{eqnarray}
f_B (s) &=& \int_0^{\infty} \frac{dy}{\pi} \ln \coth \frac{|y^2 -s|}{2} +
\sqrt{|s|}
\theta(-s)
\nonumber \\
\ln g_B (s) &=& - 2 \int_{s}^{\infty} dy \left(
\frac{df_B(y)}{dy} \right)^2  + \ln {\cal C}_2,
\label{bosecross2}
\end{eqnarray}
where ${\cal C}_2 = \pi^{1/2} e^{1/2} 2^{-1/3} A_G^{-6}$,
and $-\infty < (s = \mu / k_B T) < \infty$.
As in the Ising case, both
crossover functions are analytic functions of $s$,
for all $-\infty
< s <
\infty$. In particular, and despite appearances, the result for $f_B (s)$ in
(\ref{bosecross2})
is analytic at $s=0$.
Refs~\cite{korepbook,koreppapers} give different expressions for the crossover
functions in the region $s<0$, whose equivalence to (\ref{bosecross2}) has not
so far
been explicitly established---we provide the missing proof in
Appendix~\ref{boseproof}.
The proof requires a new identity involving Glaisher's constant very similar
to, but
distinct from, that obtained in Section~\ref{statics}:
\begin{equation}
\ln \left( \pi^{1/2} e^{1/2} 2^{2/3} A_G^{-6} \right) =
2 \int_{-\infty}^{-1} dy \left[ \left( \frac{df_B (y)}{dy} \right)^2 +
\frac{1}{4
y} \right] + 2 \int_{-1}^{\infty} dy \left( \frac{df_B (y)}{dy} \right)^2 .
\end{equation}
Again, we do not have an analytical proof of this identity, but have obtained
overwhelming numerical evidence for its validity (Appendix~\ref{boseproof}).

The striking formal similarity between the results (\ref{bosecross2}) for the
quantum
transition in  the dilute Bose gas, and the results (\ref{isingcross2}) for
Ising model
should be  readily apparent. Transforming to dimensionful quantities, the main
difference between the results for the correlation length is that the
non-relativistic
dispersion $\hbar^2 k^2 /(2 m_B) - \mu$ of the $z=2$ Bose gas has been
replaced by the relativistic $\left( m^2 c^4 + \hbar^2 c^2 k^2 \right)^{1/2}$
in the
$z=1$ Ising model; otherwise the functional forms are essentially identical.
The
expression for $g_B$ in terms of $f_B$ is also similar to that for $g_I$ in
terms of
$f_I$; the latter expression contains an additional integral $\int dy/4y$,
which is
surely related to the presence of anomalous dimensions in the Ising model.
Note,
however, that the asymptotic behaviors of
$f_I$, $g_I$ are quite different from those of $f_B$, $g_B$.

To emphasize the significance of the above similarity, we note that it  is
present
despite some important differences between the Bose gas and the quantum Ising
model.
Both models are `solved' by mapping to models of free fermions---however, the
Bose
field $\Psi_B$ only has non-zero matrix elements between states whose fermion
number
differ by unity; in contrast, the Ising spin
$\sigma_z$ has non-zero matrix elements between states with arbitrary  numbers
of
fermions. Second, the $\mu >0$ ground state of the Bose gas has gapless
excitations,
while the Ising model has a gap for all $m \neq 0$.

\section{Discussion}
\label{conc}
In this paper, we have obtained universal crossover functions  (Eqn.
(\ref{isingcross2})) of static observables of the finite $T$
quantum Ising chain. The crossover functions turned out to be remarkably
similar in form to those
obtained earlier for the dilute Bose gas~\cite{korepbook,koreppapers,sss}.
This similarity suggests that the sophisticated
mathematical structure described in Ref~\cite{korepbook} to  describe
correlators of
the Bose gas, should also apply to the continuum limit of the Ising model in a
transverse field. Application of these methods to the Ising model should yield
more
information on the scaling function $\Phi_I$; in particular we should be able
to study
unequal time correlators away from the critical point. Traditional analyses of
the
Ising model have been carried out on a lattice~\cite{kogut,mccoy,barouch},  but
the
simplifications suggested here appear only in the continuum limit.
It would clearly
be preferable to use, instead, an analysis   which is valid in the continuum
limit from
the outset. Fortunately, just such an analysis has appeared in some recent
work~\cite{babelon,andre}; these papers
use non-trivial results on the form factors of the continuum Ising
model~\cite{formfac}
to deduce properties of the correlation functions. The existing results are at
$T=0$,
but the author has some preliminary results on extending the determinant
representation of
Ref~\cite{andre} to non-zero $T$~\cite{unpub}. Application of the methods
of Ref~\cite{korepbook} to this determinant representation is the next step,
but has not been carried out yet.

Such an analysis should be able to test the conjectured  RC scaling forms of
Section~\ref{dynamics} by a direct
solution of the quantum problem. If achieved, such
a result would yield the exact value of the dimensionless `rate constant'
$\gamma$.
Further, it would make a connection between two rather different approaches to
dynamic
phenomena---the quantum dynamics couples the Ising model to a heat bath only to
set the initial
density matrix
$\exp(-{\cal H}/{k_B T})$, and the subsequent evolution  is completely unitary;
in
contrast, the classical Glauber dynamics has the spins in constant contact with
a heat
bath.  Notice that the classical Glauber dynamics has purely relaxational
behavior for
the $\sigma_z$ spins; such a behavior might be questioned on the grounds that
$H_I$ is
integrable and possesses an infinite number of conservation laws. However, the
conservation laws are expressed simply using the fermionic degrees of freedom,
and are
highly non-local in terms of the $\sigma_z$ variables; as a result, we believe
that the
integrability will not disrupt the expected mapping to the relaxational Glauber
dynamics; for a related discussion on the role of integrability, see
Ref~\cite{jensen}.  In any event, it is clear that a direct quantum study of
the
dynamics away from criticality will surely enhance our understanding of
dissipative
dynamics in macroscopic quantum systems.

\acknowledgements
I am grateful to B.M. McCoy and J.H.H. Perk for their interest,  for clarifying
many
aspects of their results, and for a number of valuable comments on the
manuscript. I
thank V.E. Korepin, A. LeClair, S. Majumdar, N. Read, R. Shankar, and R.E.
Shrock for
helpful discussions. The research was supported by NSF Grant No.\ DMR-92-24290.

\appendix
\section{Derivation of crossover functions of the Ising model}
\label{cross}
It appears worthwhile to sketch a derivation from first principles  using a
consistent
notation.

First, following Lieb, Schultz and Mattis~\cite{kogut}, convert $H_I$ into a
free fermion
Hamiltonian by the Jordan-Wigner transformation. Then, evaluate the equal-time
spin correlator in
terms of the free-fermion correlators. This yields an expression for the
correlator in terms
of a Toeplitz determinant~\cite{kogut,barouch}:
\begin{equation}
\left\langle\sigma_z (i) \sigma_z (i+n) \right\rangle =
\left|
\begin{array}{cccc}
G_0 & G_{-1} & \cdot\cdot\cdot & G_{-n+1} \\
G_1 & & & \\
\cdot & & & \\
\cdot & & & \\
\cdot & & G_0 & G_{-1} \\
G_{n-1} & & G_1 & G_0
\end{array} \right|
\label{toeplitz}
\end{equation}
where
\begin{equation}
G_p = \int_0^{2 \pi} \frac{d\phi}{2 \pi} e^{-ip\phi} \tilde{G} (\phi)
\end{equation}
with
\begin{equation}
\tilde{G}(\phi) = \left( \frac{1-ge^{i\phi}}{1-ge^{-i\phi}} \right)^{1/2}
\tanh \left[ \frac{J}{k_B T} ((1-ge^{i\phi})(1-ge^{-i\phi}))^{1/2} \right]
\label{ggg}
\end{equation}
We can now take the large $n$ limit of (\ref{toeplitz}) by Szego's
lemma~\cite{wu} provided $g <
g_c = 1$. For $g \geq g_c$ a more complicated analysis is necessary to evaluate
(\ref{toeplitz}) directly~\cite{mccoy}; we shall not need this here as we shall
obtain results for
$g \geq g_c$ by our method of analytic continuation.
By Szego's lemma we obtain
\begin{equation}
\lim_{n \rightarrow \infty} \left\langle\sigma_z (i) \sigma_z (i+n)
\right\rangle
\sim e^{n \lambda_0} \exp \left( \sum_{p=1}^{\infty} p \lambda_p \lambda_{-p}
\right)~~,~~g < 1
\label{szego}
\end{equation}
where
\begin{equation}
\ln \tilde{G} (\phi) = \sum_{p=-\infty}^{\infty} \lambda_p e^{ip\phi}
\label{lnG}
\end{equation}

The correlation length is clearly $\xi_I = -a /\lambda_0$, and the above
approach gives us
the value of $\xi_I$ for $g < g_c$.
{}From (\ref{ggg}-\ref{lnG}) we can deduce that
\begin{equation}
\xi^{-1}_I = a \int_{-\pi}^{\pi} \frac{d \phi}{2 \pi} \ln \coth \left[
\frac{J}{k_B T}
(1 + g^2 - 2 g \cos \phi)^{1/2} \right]~~,~~g<1
\end{equation}
Now change variables of integration to $k = \phi /a$, re-express in terms of
renormalized couplings by writing $J = \hbar c/2
a$ and
$g = 1 - mca/\hbar$, and take the scaling limit $a \rightarrow 0$. This yields
\begin{equation}
\xi^{-1}_I = \int_{-\infty}^{\infty} \frac{d k}{2 \pi} \ln \coth \left[
\frac{(m^2 c^4 + \hbar^2 c^2 k^2)^{1/2}}{2 k_B T} \right]~~,~~m>0
\end{equation}
Let the scaling function for the inverse correlation
length, in (\ref{isingcross1}), $f_I (s) =  f_I^{>} (s)$ for $s > 0$, where
recall that
$s = mc^2 / k_B T$. Then clearly
\begin{equation}
f_I^> (s) = \frac{1}{\pi} \int_{0}^{\infty} dy \ln \coth \frac{(y^2 +
s^2)^{1/2}}{2}.
\label{fi2}
\end{equation}
Notice that $f_I^> (s)$ is an even function of $s$.
Later, it will be useful to have another form for $f_I^> (s)$: change
variables of
integration
$y \rightarrow (y^2 - s^2)^{1/2}$, and integrate by parts to obtain
\begin{equation}
f_I^> (s) = \frac{1}{\pi} \int_{|s|}^{\infty} \frac{dy}{\sinh y} (y^2 -
s^2)^{1/2}
\label{fi1}
\end{equation}

We now need to get the scaling function of the inverse correlation  length,
$f_I (s)$
for $s \leq 0$. We could get this by returning to an analysis of
(\ref{toeplitz}), but
as noted earlier, Szego's lemma does not directly apply and the analysis is not
straightforward~\cite{mccoy}. Instead, we use the method of analytic
continuation to
obtain the answer quite rapidly. Unlike the needed $f_I (s)$, the function
$f_I^> (s)$ is not analytic at $s=0$. Let us  examine the nature of the
singularity
in $f_I^> (s)$ near $s=0$.
Insert $1 = ((y^2 + s^2)/(y^2+1))^{1/2} ((y^2 + 1)/(y^2+s^2))^{1/2}$ in
the logarithm in (\ref{fi2}) to get
\begin{equation}
f_I^> (s) = \frac{1}{\pi} \int_{0}^{\infty} dy \ln \left[
\left(\frac{y^2+s^2}{y^2+1}\right)^{1/2} \coth \frac{(y^2 + s^2)^{1/2}}{2}
\right]
+ \frac{1}{2\pi} \int_0^{\infty} dy \ln \left(\frac{y^2+1}{y^2+s^2}\right)
\end{equation}
The first integral has an integrand which is a smooth function of $s^2$ for
all values
of $y$ ($x \coth x$ is regular at $x=0$ and has a series expansion with only
even
powers of
$x$); so for small
$s$ it can only yield a power series containing
 even, non-negative powers of $s$. The second integral can be performed exactly
 and we
obtain
\begin{equation}
f_I^> (s \rightarrow 0)
= - |s|/2 + \mbox{terms with even, non-negative powers of $s$}
\end{equation}
The $-|s|/2$ term is the only non-analyticity at $s=0$. It is then evident that
defining
\begin{equation}
f_I (s) = f_I^> (s) + |s| \theta (-s)
\end{equation}
gives the unique analytic function
which equals $f_I^> (s)$ for $s>0$; this was the result quoted in
(\ref{isingcross2})
for the crossover function of the correlation length, and is valid for all
values of
$s$.

We turn next to the scaling function, $g_I (s)$, of the amplitude. We need to
evaluate the
summation in (\ref{szego}); techniques for doing this were developed in
Ref~\cite{mccoy} and we
shall not reproduce them here.
The result is contained in
Eqn. (3.27)~\cite{caution} of Ref~\cite{mccoy}, and taking its scaling limit in
a manner similar
to that discussed above for the correlation length, we obtain
\begin{equation}
\ln g_I (s) = \frac{1}{2 \pi^2} \int_{s}^{\infty} \frac{dy}{\sinh y}
\int_{s}^{\infty}
\frac{d\tilde{y}}{\sinh \tilde{y}} \ln \left| \frac{(y^2 -s^2)^{1/2}+
(\tilde{y}^{2}
-s^2)^{1/2}}{(y^2 -s^2)^{1/2}- (\tilde{y}^{2}
-s^2)^{1/2}} \right| + \frac{1}{4} \ln s~~,~~s>0.
\label{gi1}
\end{equation}
We now have to analytically continue
this result to obtain the scaling function for all values of
$s$. We first take its derivative and obtain
\begin{equation}
\frac{1}{g_I (s)} \frac{dg_I (s)}{ds} = - \frac{1}{\pi^2} \int_{s}^{\infty}
\frac{dy}{\sinh y} \int_{s}^{\infty}
\frac{d\tilde{y}}{\sinh \tilde{y}}~
\frac{s}{((y^2 -s^2)(\tilde{y}^{2}
-s^2))^{1/2}}  + \frac{1}{4s} ~~,~~s>0.
\end{equation}
Remarkably, the integrals over $y$ and $\tilde{y}$ have factorized. Now
notice by comparing with (\ref{fi1}) that
\begin{equation}
\frac{1}{g_I (s)} \frac{dg_I (s)}{ds} = - \frac{1}{s} \left( \frac{df_I^>
(s)}{ds} \right)^2
+ \frac{1}{4s}~~,~~s>0
\label{gi2}
\end{equation}
{}From (\ref{gi1}) it is evident that $\ln g_I (s \rightarrow +\infty) =
(1/4)\ln s$; using this
boundary condition we may integrate (\ref{gi2}) to obtain
\begin{equation}
g_I (s) =  \exp\left[  \int_{s}^{\infty} \frac{dy}{y} \left(
\frac{df_I^>(y)}{dy} \right)^2 - \int_s^{1} \frac{dy}{4y} \right]~~,~~s>0
\end{equation}
Because $df_I^> /dy |_{y\searrow 0} = -1/2$, the above integral is a smooth
function of
$s$ as $s\searrow 0$.
The result (\ref{isingcross2}) for $g_I (s)$, for all $s$,
now follows simply by analytically continuing $f_I^> (y)$ to $f_I (y)$.

Finally, we consider the asymptotic limits in (\ref{asymptote2}). Those for
$f_I (s)$
can be obtained by elementary methods. The limit $g_I (s \rightarrow +\infty)$
is also
evident from  (\ref{isingcross2}). Consider next $g_I (s \rightarrow -\infty)$.
Use
(\ref{isingcross2}), and express $f_I (s)$ in terms of $f_I^> (s)$; after some
elementary manipulations we obtain
\begin{equation}
\ln g_I (s \rightarrow -\infty) = -\frac{3}{4} \ln |s| - \int_0^1 \frac{dy}{y}
\left(2 \frac{df_I^> (y)}{dy} + 1 \right) - 2 \int_1^{\infty} \frac{dy}{y}
\frac{df_I^> (y)}{dy}
\label{gi3}
\end{equation}
After using (\ref{fi2}), and interchanging orders of integration, all integrals
can  be found in
standard tables.  The integrals in (\ref{gi3}) actually sum to 0, and we obtain
the
result in (\ref{asymptote2}): $\ln g_I (s \rightarrow -\infty) = -(3/4) \ln
|s|$.

Finally, consider the value of $g_I (0)$. We have been unable to evaluate the
integrals in
(\ref{isingcross2}) analytically. Instead, we carried out a high precision
numerical
integration, using the arbitrary precision numerics built into the {\sc
Mathematica}
package. This gave us
\begin{equation}
\ln g_I (0) = -0.152318694592340635
\label{gi6}
\end{equation}
Compare this with the numerical value of the expression obtained by combining
the $T=0$ results of
Pfeuty with conformal invariance:
\begin{equation}
\ln \left( \pi^{1/4} e^{1/4} 2^{1/12} A_G^{-3} \right) =
-0.152318694592340634987799466\ldots.
\label{gi5}
\end{equation}
There is  essentially no doubt that the expression in (\ref{gi5}) is the exact
value of
$\ln g_I (0)$, and this gives us one the claimed representations of Glaisher's
constant.

\section{Scaling limit of Glauber's solution}
\label{glauberscale}
I thank S. Majumdar for showing me how to take the scaling limit discussed in
this
Appendix.

Glauber~\cite{glauber} considered the classical Ising model (Eqn
(\ref{hamising}) at $g=0$) coupled
to a heat bath at a temperature $T$. He derived a differential equation for the
correlator
$C (i, t) = \left\langle \sigma_z (i,t) \sigma_z (0,0) \right\rangle$:
\begin{equation}
\frac{1}{\alpha} \frac{\partial C(i,t)}{\partial t} = - C(i,t) + \frac{1}{2}
\tanh \frac{2J}{k_B T} \left[ C(i-1,t) + C(i+1,t) \right]
\label{glauber1}
\end{equation}
where $\alpha$ is a relaxation rate constant. At low $T$, the classical Ising
model has
a correlation length $\xi = (a/2) e^{2J/k_B T}$ ($a$ is the lattice spacing)
and the
equal time correlator obeys $C(x=ia, 0) = e^{-|x|/\xi}$. As $\xi \gg a$, we may
take the continuum
limit of (\ref{glauber1}) and obtain the partial differential equation
\begin{equation}
\frac{1}{\alpha a^2} \frac{\partial C(x,t)}{\partial t} = \frac{1}{2}
\frac{\partial^2
C(x,t)}{\partial x^2} - \frac{C(x,t)}{2 \xi^2}
\end{equation}
This equation is to be solved subject to the initial condition
$C(x,0)=e^{-|x|/\xi}$.
This is easily done by a spatial Fourier transform and we obtain the result in
(\ref{RCscale},\ref{erfc}), with the unknown constant $\gamma$  proportional to
$\alpha
a^2$.

\section{Computations for the Bose gas}
\label{boseproof}

In this appendix we will establish the equivalence of our results for  the
crossover
functions of the dilute Bose gas, $f_B (s)$ and $g_B (s)$ in
(\ref{bosecross2}), to
those for $s <0$ in Refs~\cite{koreppapers}.

As in the case of the Ising model in Appendix~\ref{cross}, it is useful to
define the function
$f_B^> (s)$,
\begin{equation}
f_B^> (s) = \int_0^{\infty} \frac{dy}{\pi} \ln \coth \frac{|y^2 - s|}{2},
\end{equation}
such that $f_B (s) = f_B^> (s) + \sqrt{-s} \theta (-s)$.
The proof that $f_B (s)$ is analytic at $s=0$ is very similar to that presented
in
Appendix~\ref{cross} for $f_I (s)$, and will be omitted.

Let us call $g_{BK}$, the result for the amplitude $g_B (s)$  obtained in
Ref~\cite{koreppapers} for $s < 0$. Then~\cite{koreppapers}
\begin{equation}
g_{BK} (s) = \frac{1}{\sqrt{-4 s}} \exp \left( -
\frac{2\sqrt{-s}}{\pi} \int_0^{\infty} \frac{dy}{y^2 - s} \ln \coth \frac{y^2 -
s}{2}
+ 2 \int_{-\infty}^{s} dy \left( \frac{d f_B^>}{dy} \right)^2 \right)~~,~~s<0
\end{equation}
Clearly we have $\lim_{s\rightarrow -\infty} \sqrt{-4s} g_{BK} (s) = 1$.
For $\lim_{s\rightarrow -\infty} \sqrt{-4s} g_{B} (s) $ to equal unity, we must
have
from (\ref{bosecross2}) that
the universal number
\begin{equation}
{\cal C}_3 = 2 \int_{-\infty}^{-1} dy  \left[ \left( \frac{df_B (y)}{dy}
\right)^2 +
\frac{1}{4 y} \right] + 2 \int_{-1}^{\infty} dy \left( \frac{df_B (y)}{dy}
\right)^2
\label{gb1}
\end{equation}
is equal to
\begin{equation}
\ln \left( \pi^{1/2} e^{1/2} 2^{2/3} A_G^{-6} \right) =
0.04193620109529\ldots
\label{gb2}
\end{equation}
We evaluated (\ref{gb1}) numerically and found
\begin{equation}
{\cal C}_3 = 0.0419362011
\end{equation}
which is in complete agreement with (\ref{gb2}).  The equations
(\ref{gb1},\ref{gb2})
therefore contain one of the claimed integral representations of Glaisher's
constant,
$A_G$. The integrand in
(\ref{gb1}) only has a power-law convergence at
$+\infty$, and hence the numerical integration is  not as accurate as that
leading to
(\ref{gi6}), which converged exponentially fast. This difference between  the
two
representations is directly linked to the fact that the Ising model has a gap
at
$T=0$ for all $m\neq 0$, while the Bose gas has gapless excitations for $\mu >
0$.

Let us now assume that $\lim_{s\rightarrow -\infty} \sqrt{-4s} g_{B} (s) = 1$.
Then proving the equality between $g_B (s)$ and $g_{BK} (s)$ for all  $s < 0$
can be
shown (after substituting $f_B (s) = f_B^> (s) + \theta(-s)
\sqrt{-s}$ in (\ref{bosecross2}) and performing some elementary manipulations)
to be
equivalent to establishing that
\begin{equation}
\int_{-\infty}^{s} \frac{dy}{\sqrt{-y}} \frac{d f_B^>}{dy}
= \frac{\sqrt{-s}}{\pi} \int_0^{\infty} \frac{dy}{y^2 - s}  \ln \coth \frac{y^2
-
s}{2}~~,~~ s<0.
\label{final}
\end{equation}
As both sides of the equation vanish as $s \rightarrow -\infty$,  it is
sufficient to
establish the equality of their derivatives with respect to $s$. Taking the
derivative,
and then converting all of the $\ln \coth$ terms to $1/\sinh$ by integrating by
parts,
one can show after some straightforward, but tedious, manipulations that
(\ref{final})
is indeed true.

\newpage
\newpage
\newpage


\begin{references}
\bibitem{caution} There appears to be a crucial missing  prefactor of 1/4 in
Eqn (3.28)
and subsequent equations of Ref~\protect\cite{mccoy}. If omitted, the scaling
structure
of the results would be spoiled.
\end{references}
\end{document}